\documentclass[reqno]{amsart}
\usepackage{newlfont}
\usepackage{url}
\usepackage{epsfig,graphics}
\usepackage{amsmath,amssymb,amsthm,xspace,amscd}
\usepackage{verbatim}
\usepackage{multirow}

\newtheorem{thm}{Theorem}[section]
\newtheorem{prop}[thm]{Proposition}

\theoremstyle{definition}

\newtheorem{example}{Example}

\def\R{\mathbb R}

\def\N{\mathbb N}

\def\r{\rangle}
\def\l{\langle}

\def\l{\langle}
\def\r{\rangle}

\def\l{\langle}
\def\r{\rangle}

\allowdisplaybreaks

\author{F.W.~Lemire$^1$}
\author{J.~Patera$^{2,3}$}
\author{M.~Szajewska$^{4,2}$}

\address{
$^1$ Department of Mathematics and Statistics, University of Windsor, 401 Sunset Ave., Windsor, N9B 3P4, Ontario, Canada. \\
$^2$ Centre de recherches math\'ematiques, Universit\'e de Montr\'eal, C.~P.~6128 -- Centre ville, Montr\'eal, H3C\,3J7, Qu\'ebec, Canada.\\
$^3$MIND Research Institute, 111 Academy Dr., Irvine, California 92617 
\\
$^4$ Institute of Mathematics, University of Bialystok, Akademicka~2, PL-15-267, Bialystok, Poland.}

\email{lemire@uwindsor.ca, patera@crm.umontreal.ca, m.szajewska@math.uwb.edu.pl}


\begin{document}

\title[Dominant weight multiplicities]
{Dominant weight multiplicities in hybrid characters of
$B_n$, $C_n$, $F_4$, $G_2$}



 \begin{abstract}\
The characters of irreducible finite dimensional representations of compact simple Lie group $G$ are invariant with respect to the action of the Weyl group $W(G)$ of $G$. The defining property of the new character-like functions (`hybrid characters') is the fact that $W(G)$ acts differently on the character term corresponding to the long roots than on  those corresponding to the short roots. Therefore the hybrid characters are defined for the simple Lie groups with two different lengths of their roots. Dominant weight multiplicities for the hybrid characters are determined. The formulas for `hybrid dimensions' are also found for all cases as the zero degree term in power expansion of the `hybrid characters'.
\end{abstract}

\maketitle


\section{Introduction}

Motivation to study properties of characters of irreducible representations of simple Lie groups was justified by their fundamental importance for the representation theory \cite{B,S}. Subsequently one may find the motivation in other properties of the characters, such as the possibility to discretize them uniformly for simple Lie groups of all types and ranks \cite{MP84,MP87} and, more generally, the duality between the representation theory and the conjugacy classes of elements of finite  order \cite{MP84,MP87} in compact simple Lie groups.

In this paper we consider the `hybrid' characters which are defined for simple Lie groups with two lengths of simple roots, that is the groups of  types $B_n$, $C_n$, $G_2$, and $F_4$. The hybrid characters are character-like functions. Although no related algebraic structure resembling the simple Lie groups can be attached to them. Nevertheless their existence carries valuable properties that in the long run will undoubtedly find their exploitation. Let us single out the following ones:
\newline
$\bullet$ 
New vast class of special functions and  polynomials related to characters  can be discretized. With a twist of familiar definition of degree of multivariable polynomial, their nodes and the cubature formulas appear naturally and are optimal (called Gaussian) in their efficiency \cite{X95,MMotP1,MP11}.
\newline
$\bullet$ 
New families of special functions that are orthogonal on sets of lattice points  of any density in $F_M$, and thus can be used for new type of Fourier expansions of digital data on $F_M$.


The characters and $C$-functions, that are invariant with respect to the Weyl group $W$ of the underlying semisimple Lie group. as well as the 
$S$-functions that are skew invariant with respect to $W$, and the $S^S$- and $S^C$-functions that are half invariant and half skew invariant with respect to $W$, are related also to systems of orthogonal polynomials. The simplest example are the Chebysheff polynomials of the first and second kind in one variable. On the side of greater generality related to characters and hybrid characters are the Macdonald and Jacobi polynomials \cite{Mac}. However as often is the case, more restricted systems may have useful properties that are not available in the most general case. 


Similarly as characters of finite irreducible representations can be expressed as finite sums of Weyl group invariant functions (`orbit functions of type $C$'), also the hybrid characters can be written in the the similar way. In the case of characters the coefficients of such sums are the all important dominant weight multiplicities \cite{BMP, Freud, MP11}. In case of hybrid characters, the coefficients are for the first time determined here.
 
  Corresponding Fourier transforms of digital data are yet to be exploited. The hybrid characters of this paper lead to new classes of special functions that are orthogonal on lattices of all dimensions and symmetry types. For that to be used, one needs the dominant weight multiplicities of the hybrid characters for every representation that comes into considerations.

  The later motivation gains in importance with ever increasing number of digital data that require processing and that are collected today on lattices of various densities and all possible symmetries in dimensions $\geq2$.

In general, the character $\chi_\lambda(\mathfrak g)$ of a finite dimensional representation of a compact semisimple Lie group  is a finite sum of exponential functions with exponents depending through the scalar product $\l\lambda,x\r$  on the points $x$ of the maximal torus of the Lie group and on a weight  $\lambda$ of the weight system of the representation. For many years a practical obstacle in exploiting the characters of other than a few lowest representations, was the relative inefficiency of the general algorithm \cite{Freud} for computing the multiplicities of weights $\lambda$ of the representation. This obstacle was removed by the fast algorithm \cite{MP82,BMP}. Truly large scale applications involving the characters became then possible (for example \cite{GP,MNP}).

In this paper the problem of multiplicities of weights is solved for the recently discovered  character-like functions (denoted by $\chi^L_\lambda(\mathfrak g)$  and $\chi^S_\lambda(\mathfrak g)$), called hybrid characters \cite{MotP,Sz,MMotP1}, for any simple Lie group with roots of two different lengths (types $B_n$, $C_n$, $F_4$, and $G_2$). More precisely, the present paper solves the problem of  determining the coefficients $p^\lambda_\mu(\mathfrak g)$ and $q^\lambda_\mu(\mathfrak g)$ arising in $\chi^L_\lambda(\mathfrak g)$  and $\chi^S_\lambda(\mathfrak g)$ respectively, when the hybrid characters are written as linear combinations of the $C$-functions. Once those coefficients are known,  the discretization of the $C$-functions \cite{MP84, MP87} can be directly extended to the hybrid characters. 

Main result of the paper is the demonstration of the fact that the weight multiplicities for the hybrid characters of a group $G$ are found using the well known multiplicities of several representations of an appropriate subgroup $G'\subset G$. A possible alternate way to get the hybrid multiplicities, namely modification of the algorithms \cite{Freud, MP82}, would be more complicated.

The new classes of character-like functions are formed by defining two new homomorphisms from the Weyl group $W(\mathfrak g)$ of the simple Lie algebra $\mathfrak g$ to $\{\pm1\}$ which generalize the usual sign function.
\begin{gather}\label{auto}
\sigma^L, \sigma^S : W(\mathfrak g)\longrightarrow\{\pm1\}\,,
\end{gather}

In this paper we consider the cases of Lie groups/Lie algebras of types $B_n$, $C_n$ of any rank $n$, as well as  $F_4$ and $G_2$. The homomorphisms \eqref{auto} are defined by distinguishing their values on  the reflections with respect to hyperplanes orthogonal to long and short roots of $\mathfrak g$.
Such a possibility was first noted in \cite{MotP} in the case of $C_2$, while considering the 2-variable specialization of the $n$-variable case \cite{KPtrig}.  In \cite{MMotP1} the hybrid characters are described for all simple Lie algebras with 2-root lengths. The definitions use the homomorphisms \eqref{auto} denoted by $\sigma^L$ and $\sigma^S$, Also the $G_2$ case is found in \cite{Sz}. For every representation of those groups there are two  new character-like functions, 
$\chi^L_\lambda(\mathfrak g)$ and $\chi^S_\lambda(\mathfrak g)$, in addition to their corresponding character $\chi_\lambda(\mathfrak g)$ functions.

The role of irreducible characters in representation theory of simple Lie groups cannot be overestimated. Therefore it is interesting to investigate the structure and the role of the new `hybrid' characters that so closely resemble the well known irreducible character functions.

Interesting problem, unanswered so far,  is about the possibility to link the hybrid characters to group-like algebraic structures, where the hybrids would serve as the characters. This is a challenging question with no guarantee that a positive answer can be found to it.  Regardless of whether such a link  can be found,  their orthogonality opens new possibilities of applications in Fourier analysis. Expansion of functions on the fundamental region of compact semisimple Lie group has been shown in general \cite{KP06,KP07a,KP07b,KP08,KP08b,KP09} with  number of cases studied in detail in dimension 2 or 3 \cite{X12,X10b, HP09,HHP,HMP,HMP1,HP2,Sz}. Moreover, such expansions exist for data that are continuous functions of their variables, as well as data sampled on a lattice of any density in the fundamental region of the Lie group. Equally interesting is the existence of the Gaussian cubature formula for the characters \cite{X10a,X95,MP11}. Most recently it was shown that also the hybrid characters admit cubature relation \cite{MMotP1}, although only part of them would have the Gaussian optimality.

Among the properties of the character $\chi_\lambda(\mathfrak g)$ that are also shared with the hybrid characters $\chi^L_\lambda(\mathfrak g)$  and $\chi^S_\lambda(\mathfrak g)$, let us point out the following ones:
\smallskip

\noindent
$\bullet$\ Completeness as the bases in their respective functional spaces \cite{MMotP1}.
\newline
$\bullet$\ Orthogonality \cite{HMP2} when integrated over the fundamental region $F(\mathfrak g)$.
\newline
$\bullet$\ Decomposition of products of the functions into their sums \cite{MotP}.
\newline
$\bullet$\ Discrete orthogonality \cite{HMP2} when summed up over a fragment of a lattice $F_M(\mathfrak g)$ in $F(\mathfrak g)$, where density of the lattice is fixed by our choice of $M\in\N$.
\newline
$\bullet$\ Hybrid analogs of the well known Weyl formulas for dimensions of irreducible representations obtained by by putting the character variables equal to zero, are also found here. They only underline the question to what group-like structure they pertain?
\smallskip

Another  useful application of the hybrid characters is a consequence of their well-defined behaviour at the boundary $\partial F$ of their domain $F$. While irreducible characters and the $C$-functions are symmetric at $\partial F$, the $S$-functions are skew symmetric at $\partial F$. The hybrid characters are symmetric and skew symmetric at different parts of $\partial F$. Hence Fourier expansions  of functions with similar boundary behaviour would be most simply done in terms of expansions into series of corresponding hybrid characters.

\section{Preliminaries}

Let $\mathfrak g$ be one of the simple Lie algebras
\begin{gather}
B_n\ (n\geq2),\qquad C_n\ (n\geq2),\qquad F_4,\qquad G_2\,.
\end{gather}
The Lie algebras $B_2$ and $C_2$ are isomorphic but it is convenient to keep both of them. We adopt the Dynkin numbering of simple roots, see for example \cite{BMP}.

In this section we establish a convenient notation for the root systems and Weyl groups for each of these algebras that we will use in subsequent sections. Let $\{e_1,\dots, e_n\}$ denote the orthonormal basis
($e$-basis) of the real Euclidean space $\R^n$ of dimension $n$. We express in the $e$-basis expansions for a base of simple roots ($\alpha$-basis), and the corresponding fundamental weights ($\omega$-basis) of $\mathfrak g$. In particular, we know the Cartan matrix $\mathfrak C$ for any $\mathfrak g$.
$$
\mathfrak C=\left(2\frac{\l\alpha_j\mid\alpha_k\r}{\l\alpha_k\mid\alpha_k\r}\right)\,,
\qquad j,k=1,2,\dots,n\,.
$$
where $\langle\cdot\mid\cdot\rangle$ is the usual scalar product in $\R^n$. These are well known \cite{B,BMP} standard attributes of any simple $\mathfrak g$. The matrix $\mathfrak C$ links the $\alpha$- and $\omega$-bases,
$$
\alpha=\mathfrak C\omega\,,\qquad \omega=\mathfrak C^{-1}\alpha.
$$

A reflection $r_\xi$ with respect to a hyperplane orthogonal to $\xi$ and containing the origin in $\R^n$, is given by
\begin{gather}
r_\xi x=x-\frac{2\langle x\mid\xi\rangle}{\langle \xi\mid\xi\rangle}\xi\,,
\qquad x,\xi\in\R^n\,.
\end{gather}
The Weyl group of $\mathfrak g$, denoted W($\mathfrak g$), is the reflection group generated by $\{r_{\alpha_i}\}$.
\subsection{The Lie algebra $B_n$}\

The $e$-basis realization of a base of  simple roots and the corresponding fundamental weights of $B_n$  is given by

$$\begin{matrix} \hbox{simple roots}\\ \alpha_1=e_1-e_2\\ \alpha_2=e_2-e_3\\ \vdots \\ \alpha_{n-1}=e_{n-1}-e_n\\ \alpha_n=e_n \end{matrix}  \hskip .5 in  \begin{matrix} \hbox{fundamental weights}\\ \omega_1=e_1\\ \omega_2=e_1+e_2\\ \vdots \\ \omega_{n-1}=e_1+\cdots +e_{n-1}\\ \omega_n=\frac{1}{2}(e_1+\cdots +e_n) \end{matrix}$$

Let $\rho^L$ (respectively $\rho^S$) denote half the sum of the long (respectively short) positive roots of $B_n$. Then we have
\begin{equation}
\rho^L =\omega_1+\ldots +\omega_{n-1}, \qquad \quad  \rho^S =\omega_n.
\end{equation}

Recall that the Weyl group $W(B_n)$ of $B_n$ can be viewed as the set of all permutations and sign changes on the subscripts $\{1,\dots, n\}$ of the $e$-basis. With this identification, if $\sigma^L$ and $\sigma^S$ denote the long and short sign homomorphisms from $W(B_n)$ to $\{\pm 1 \}$, we have:

$$ \begin{matrix} r_{\alpha_1}=(1,2)\\ r_{\alpha_2}=(2,3)\\ \vdots \\ r_{\alpha_{n-1}}=(n-1,n)\\ r_{\alpha_n}=n\to -n\end{matrix}\hskip .5 in \begin{matrix} \sigma^L(r_{\alpha_1})=-1\\ \sigma^L(r_{\alpha_2})=-1\\ \vdots\\ \sigma^L(r_{\alpha_{n-1}})=-1\\
 \sigma^L(r_{\alpha_n})=+1\end{matrix}\hskip .5 in \begin{matrix} \sigma^S(r_{\alpha_1})=+1\\ \sigma^S(r_{\alpha_2})=+1\\ \vdots\\ \sigma^S(r_{\alpha_{n-1}})=+1\\ \sigma^S(r_{\alpha_n})=-1\end{matrix}$$

The set of long roots in the root system of  $B_n$ is a root system equivalent to the root system of $D_n$. The set of short roots of the root system of $B_n$ is a root system equivalent to the root system of $nA_1:=A_1\oplus \cdots \oplus A_1$.

\subsection{The Lie algebra $C_n$}\

The $e$-basis realization of a base of  simple roots and the corresponding fundamental weights of $C_n$  is given by

$$\begin{matrix} \hbox{simple roots}\\ \alpha_1=e_1-e_2\\ \alpha_2=e_2-e_3\\ \vdots \\ \alpha_{n-1}=e_{n-1}-e_n\\ \alpha_n=2e_n \end{matrix}  \hskip .5 in  \begin{matrix} \hbox{fundamental weights}\\ \omega_1=e_1\\ \omega_2=e_1+e_2\\ \vdots \\ \omega_{n-1}=e_1+\cdots +e_{n-1}\\ \omega_n=e_1+\cdots +e_n \end{matrix}$$

Let $\rho^L$ (respectively $\rho^S$) denote  half sum of the long (respectively short) positive roots of $C_n$. Then we have
\begin{equation}
\rho^L =\omega_n, \qquad \quad  \rho^S =\omega_1+\cdots +\omega_{n-1}.
\end{equation}

Recall that the Weyl group $W(C_n)$ of $C_n$ can be viewed as the set of all permutations and sign changes on the subscripts $\{1,\dots, n\}$ of the $e$-basis. With this identification, if $\sigma^L$ and $\sigma^S$ denote the long and short sign homomorphisms from $W(C_n)$ to $\{\pm 1 \}$, we have:

$$ \begin{matrix} r_{\alpha_1}=(1,2)\\ r_{\alpha_2}=(2,3)\\ \vdots \\ r_{\alpha_{n-1}}=(n-1,n)\\ r_{\alpha_n}=n\to -n\end{matrix}\hskip .5 in \begin{matrix} \sigma^L(r_{\alpha_1})=+1\\ \sigma^L(r_{\alpha_2})=+1\\ \vdots\\ \sigma^L(r_{\alpha_{n-1}})=+1\\
 \sigma^L(r_{\alpha_n})=-1\end{matrix}\hskip .5 in \begin{matrix} \sigma^S(r_{\alpha_1})=-1\\ \sigma^S(r_{\alpha_2})=-1\\ \vdots\\ \sigma^S(r_{\alpha_{n-1}})=-1\\ \sigma^S(r_{\alpha_n})=+1\end{matrix}$$

The set of long roots in the root system of  $C_n$ is a root system equivalent to the root system of $nA_n$. The set of short roots of the root system of $C_n$ is a root system equivalent to the root system of $D_n$.

\subsection{The Lie algebra $F_4$}\

The $e$-basis realization of a base of  simple roots and the corresponding fundamental weights of $F_4$  is given by

$$\begin{matrix} \hbox{simple roots}\\ \alpha=e_2-e_3\\ \beta=e_3-e_4\\ \gamma=e_4\\ \delta=\frac{1}{2}(e_1-e_2-e_3-e_4) \end{matrix}  \hskip .5 in \ \begin{matrix} \hbox{fundamental weights}\\ \omega_1=e_1+e_2\\ \omega_2=2e_1+e_2+e_3 \\ \omega_3= \frac{1}{2}(3e_1+e_2+e_3+e_4)\\ \omega_4=e_1\end{matrix}$$

Let $\rho^L$ (respectively $\rho^S)$ denote one half the sum of the long (respectively short) positive roots of $F_4$, then we have
$$\rho^L=\omega_1+\omega_2\ \hbox{ and } \ \rho^S=\omega_3+\omega_4.$$

The Weyl group W($F_4$) of $F_4$ can be viewed as the set of all permutations and sign changes on the subscripts $\{1,2,3,4\}$ together with the maps sending the basis $\{e_1, \dots, e_4\}$  to an orthogonal basis of vectors in $\{\frac{1}{2}(\pm e_1\pm e_2
\pm   e_3\pm e_4)\}$.

If $\sigma^L$ and $\sigma^S$ denote the long and short sign homomorphisms from W($F_4$) to $\{\pm 1 \}$ we have:

$$ \begin{matrix} r_{\alpha}=(2,3)\\ r_\beta=(3,4)\\ r_\gamma= 4\to -4\\ r_\delta=\left\{ \begin{matrix}e_1\to \frac{1}{2}(e_1+e_2+e_3+e_4)\\ e_2\to \frac{1}{2}(e_1+e_2-e_3-e_4)\\ e_3\to \frac{1}{2}(e_1-e_2+e_3-e_4)\\ e_4\to \frac{1}{2}(e_1-e_2-e_3+e_4)\end{matrix}\right.\end{matrix}\hskip .5 in \begin{matrix} \sigma^L(r_{\alpha})=-1\\ \sigma^L(r_\beta)=-1\\ \sigma^L(r_\gamma)=+1\\ \sigma^L(r_\delta)=+1\end{matrix}\hskip .5 in \begin{matrix} \sigma^S(r_{\alpha})=+1\\ \sigma^S(r_\beta)=+1\\ \sigma^S(r_\gamma)=-1\\ \sigma^S(r_\delta)=-1\end{matrix}$$

The set of long roots in the root system of  $F_4$ is a root system equivalent to the root system of $D_4$. The set of short roots of the root system of $F_4$ is a root system equivalent to the root system of $D_4$.

\subsection{The Lie algebra $G_2$}\

The $e$-basis realization of a base of  simple roots and the corresponding fundamental weights of $G_2$  is given by

$$\begin{matrix} \hbox{simple roots}\\ \alpha=e_1-e_2\\ \beta=\frac{1}{3}(-e_1+2e_2-e_3) \end{matrix}  \hskip .5 in  \begin{matrix} \hbox{fundamental weights}\\ \omega_1=e_1-e_3\\ \omega_2=\frac{1}{3}(e_1+e_2-2e_3) \end{matrix}$$

Let $\rho^L$ (respectively $\rho^S$) denote one half the sum of the long (respectively short) positive roots of $G_2$, then we have
$$ \rho^L=\omega_1\hbox{ and } \rho^S=\omega_2.$$

Recall that the Weyl group W($G_2$) of $G_2$ is generated by the set of all permutations and minus the identity on the subscripts $\{1,2,3\}$. With this identification if $\sigma$ denotes the usual sign function on W($G_2$) and  $\sigma^L$,  $\sigma^S$ denote the long and short sign homomorphisms from W($G_2$) to $\{\pm 1 \}$ we have:

$$ \begin{matrix} r_{\alpha}=(1,2)\\ r_\beta=(-id)\circ(1,3)\end{matrix}\hskip .5 in \begin{matrix} \sigma(r_{\alpha})=-1\\ \sigma(r_\beta)=-1\end{matrix}\hskip .5 in \begin{matrix} \sigma^L(r_{\alpha})=+1\\ \sigma^L(r_\beta)=-1\end{matrix}\hskip .5 in \begin{matrix} \sigma^S(r_{\alpha})=-1\\ \sigma^S(r_\beta)=+1\end{matrix}$$

The set of long roots in the root system of  $G_2$ is a root system equivalent to the root system of $A_2$. The set of short roots of the root system of $G_2$ is a root system equivalent to the root system of $A_2$.

\section{Hybrid character functions}

Let $P^+(\mathfrak g)$ denote the dominant integral weights of
$\mathfrak g$, then for any $\lambda\in P^+(\mathfrak g)$ \cite{KP06,HP09}, the $W$-invariant $C$-functions and the $W$-skew invariant $S$-functions are defined as follows,

\begin{align}
C_\lambda(\mathfrak g)(x)
   &= \sum_{\mu\in W(\mathfrak g)\cdot\lambda} e^{2\pi i\l\mu|x\r}
                 \label{Cfunctions}\\
S_{\rho+\lambda}(\mathfrak g)(x)
   &= \sum_{\mu\in W(\mathfrak g)\cdot(\lambda+\rho)}
                          \sigma(\mu)\,e^{2\pi i\l\mu|x\r}
                 \label{Sfunctions}
\end{align}
where $\sigma(\mu)=\sigma(\phi)$ when $\mu=\phi\cdot (\lambda+\rho)$.  The irreducible character function is then given by

$$\chi_\lambda(\mathfrak g)(x)= \frac{S_{\lambda+\rho}(\mathfrak g)(x)}{S_\rho(\mathfrak g)(x)}$$
and the Weyl character formula expresses the character function as a linear combination of $C$-functions

 \begin{gather}\label{character}
\chi_\lambda(\mathfrak g)(x)
   =\frac{S_{\rho+\lambda}(\mathfrak g)(x)}
            {S_\rho(\mathfrak g)(x)}
            =\sum_{\mu}m^\lambda_\mu(\mathfrak g) C_\mu(\mathfrak g)(x)
\end{gather}
where, for each dominant integral weight $\mu$ of $\mathfrak{g}$, $m^\lambda_\mu(\mathfrak g)$ is the dominant weight multiplicity. It denotes  the dimension of the $\mu$ weight space in the simple $\mathfrak g$ module having highest weight  $\lambda$.

The hybrid $S$-functions are defined as follows,
 \begin{align}
S^L_{\rho^L+\lambda}(\mathfrak g)(x)
   &= \sum_{\nu\in W(\mathfrak g)\cdot(\rho^L+\lambda)}
             \sigma^L(\nu)e^{2\pi i\l\nu|x\r}
                 \\
S^S_{\rho^S+\lambda}(\mathfrak g)(x)
    &= \sum_{\nu\in W(\mathfrak g)\cdot(\rho^S+\lambda)}
            \sigma^S(\nu) e^{2\pi i\l\nu|x\r}
\end{align}
with $\sigma^L(\nu) =\sigma^L(\phi)$ (respectively $\sigma^S(\nu)=\sigma^S(\phi)$) where $\nu=\phi\cdot(\rho^L+\lambda)$ (respectively $\nu=\phi\cdot(\rho^S+\lambda)$).

The corresponding hybrid character  functions are given by

\begin{align}
\chi^L_\lambda(\mathfrak g)(x)
   &=\frac{S^L_{\rho^L+\lambda}(\mathfrak g)(x)}
          {S^L_{\rho^L}(\mathfrak g)(x)}
          =\sum_{\mu}p^\lambda_\mu(\mathfrak g) C_\mu(\mathfrak g)(x)  \label{Lcharacter}
\end{align}

\begin{align}
\chi^S_\lambda(\mathfrak g)(x)
   &=\frac{S^S_{\rho^S+\lambda}(\mathfrak g)(x)}
          {S^S_{\rho^L}(\mathfrak g)(x)}
          =\sum_{\mu}q^\lambda_\mu(\mathfrak g) C_\mu(\mathfrak{g})(x)  \label{Scharacter}
\end{align}
where the summations range over the dominant integral weights $\mu$ of $\mathfrak{g}$.

While the coefficients $m^\lambda_\mu(\mathfrak g)$ in the expansion of the irreducible characters, $\chi_\lambda(\mathfrak{g})(x)$,  are calculated by the known algorithm \cite{MP82} and are tabulated in \cite{BMP}, the coefficients $p^\lambda_\mu(\mathfrak g)$ and $q^\lambda_\mu(\mathfrak g)$ are studied here for the first time. They are the dominant weight multiplicities for the hybrid characters 
$\chi^L_\lambda(\mathfrak g)(x)$ and 
$\chi^S_\lambda(\mathfrak g)(x)$, respectively.

\section{Long root Hybrid Character Functions}

In this section we express the coefficients $p^\lambda_\mu(\mathfrak g)$ of \eqref{Lcharacter} in terms of the multiplicities  $m^\lambda_\mu$ of \eqref{character}. Since $m^\lambda_\mu$ can be efficiently calculated \cite{MP82,BMP}, we have effectively determined the values of $p^\lambda_\mu(\mathfrak g)$.

Let $\mathfrak{g}$ denote one of the simple Lie algebras $B_n, C_n, F_4, G_2$. Let $\mathcal{G}_S$ denote the subgroup of the Weyl group $W(\mathfrak{g})$ generated by the reflections $r_\alpha$ where $\alpha$ ranges over the short simple roots of $\mathfrak{g}$. We now list a number of properties of  this subgroup.

\smallskip
\noindent
\textbf{1:} By the definition of the Weyl group $W(\mathfrak{g})$, we have that $\mathcal{G}_S$ is a transversal of the Weyl group $W(\Phi_L)$ of the long root subsystem $\Phi_L$ in the Weyl group $W(\mathfrak{g})$-- i.e. $W(\mathfrak{g})=\bigcup\limits_{\phi\in\mathcal{G}_S}W(\Phi_L)\cdot\phi$.

\smallskip
\noindent
\textbf{2:} By the definition of the sign homomorphism $\sigma^L$, we have that $\sigma^L(\phi)=+1$ for all $\phi\in \mathcal{G}_S$.

\smallskip
\noindent
\textbf{3:} There exists a base of simple roots $\Delta_L$ in $\Phi_L$ such that each $\phi\in \mathcal{G}_S$ permutes these simple roots. In fact we have

\begin{table}[h]
{\footnotesize
\begin{tabular}{|c|c|c|c|}
\hline
\rule{0pt}{10pt}
$\mathfrak g$ & $\Phi_L$ & $\mathcal{G}_S$ & $\Delta_L$ \\ \hline
$B_n$ & $D_n$  &$\langle r_n\rangle$  & $\{e_1-e_2,\ldots, e_{n-1}-e_n,e_{n-1}+e_n\}$
\rule{0pt}{10pt}
\\ \hline
\rule{0pt}{10pt}
$C_n$ & $nA_1$ &$\langle r_{\alpha_1},\cdots,r_{\alpha_{n-1}}\rangle$  & $\{2e_1,\cdots, 2e_n\}$
\rule{0pt}{10pt}
\\ \hline
$F_4$ & $D_4$ &$\langle r_\gamma, r_\delta\rangle$  & $\{e_1-e_2, e_2-e_3,e_3-e_4,e_3+e_4\}$  \rule{0pt}{10pt}
\\ \hline
\rule{0pt}{10pt}
$G_2$ & $A_2$ &$\langle r_\beta\rangle$  & $\{e_1-e_2,e_2-e_3\}$
\rule{0pt}{10pt}
\\ \hline
\end{tabular}
}
\end{table}

For each case let $\{\omega'_i\}$ denote the fundamental weights corresponding to the base of simple roots $\Delta_L$. In the table below we provide the expansions for dominant integral weights $\lambda$ of $\mathfrak{g}$ in terms of the $\{\omega'_i\}$ basis. In particular we  observe that each dominant integral weight of $\mathfrak{g}$ is a dominant integral weight of $\Phi_L$. In fact we have

\begin{table}[h]
{\footnotesize
\begin{tabular}{|c|c|c|c|}
\hline
\rule{0pt}{10pt}
$\mathfrak g$ & $\lambda$  & $\Phi_L$ & $\lambda$  \\ \hline
$B_n$ & $\sum\limits_{i=1}^nN_i\omega_i$ & $D_n$ & $\sum\limits_{i=1}^{n-1}N_i\omega'_i+(N_{n-1}+N_n)\omega'_n$
\rule{0pt}{10pt}
\\ \hline
\rule{0pt}{10pt}
$C_n$ & $\sum\limits_{i=1}^nN_i\omega_i$ & $nA_1$ & $\sum\limits_{i=1}^n\sum\limits_{j=i}^n N_j \omega'_i$
\rule{0pt}{10pt}
\\ \hline
$F_4$ & $\sum\limits_{i=1}^4N_i\omega_i$ & $D_4$ & $(N_2+N_3+N_4)\omega'_1+N_1\omega'_2+N_2\omega'_3+(N_2+N_3)\omega'_4$  \rule{0pt}{10pt}
\\ \hline
\rule{0pt}{10pt}
$G_2$ & $N_1\omega_1+N_2\omega_2$ & $A_2$ & $N_1\omega_1'+(N_1+N_2)\omega'_2$
\rule{0pt}{10pt}
\\ \hline
\end{tabular}
\medskip
\caption{{\footnotesize Dominant integral weight $\lambda$ in $\omega$-basis for  the Lie algebras $\mathfrak{g}$ is written in $\omega'_i$-basis for $\Phi_L$.}}}
\end{table}

\noindent
\textbf{4:} Since each $\phi\in \mathcal{G}_S$ permutes the base of simple roots $\Delta_L$, $\phi$ also permutes the associated fundamental weights $\{\omega'_i\}$ of $\Phi_L$ and hence if $\lambda$ is a dominant integral weight of $\mathfrak{g}$ then $\phi\cdot\lambda$ is also a dominant integral weight of $\Phi_L$. In particular we have that $\phi\cdot \rho^L=\rho^L$.

\smallskip

Using the properties of $\mathcal{G}_S$ listed above we have that for any dominant weight $\lambda$ of $\mathfrak{g}$

 $$C^L_\lambda(\mathfrak{g})(x)=\sum_{\mu\in\mathcal{G}_S\cdot\lambda}C_\mu(\Phi_L)(x)$$

 $$S^L_{\rho^L+\lambda}(\mathfrak{g})(x)=\sum_{\mu\in\mathcal{G}_S\cdot\lambda}S_{\rho^L+\mu}(\Phi_L)(x)$$

 and finally
 $$\chi^L_\lambda(\mathfrak{g})(x)=\frac{S^L_{\rho^L+\lambda}(\mathfrak{g})(x)}{S^L_{\rho^L}(\mathfrak{g})(x)}=\sum_{\mu\in\mathcal{G}_S\cdot\lambda}\chi_\mu(\Phi_L)(x)$$

 Since $\chi_\mu(\Phi_L)(x)=\sum\limits_\nu m^\mu_\nu(\Phi_L)C_\nu(\Phi_L)(x)$ where $m^\mu_\nu(\Phi_L)$ denotes the multiplicity of the dominant integral weight $\nu$ in the simple module of the Lie algebra associated with $\Phi_L$ having highest weight $\mu$. Using this we conclude that

\begin{prop}\

Multiplicity $p^\lambda_\nu(\mathfrak{g})$ of the dominant integral weight $\nu$ in the simple module of the Lie algebra associated with $\Phi_L$ having highest weight $\mu$ for hybrid character $\chi^L_\lambda(\mathfrak g)(x)$ defined by formula \eqref{Lcharacter} has a  following form
 \begin{equation}
 p^\lambda_\nu(\mathfrak{g})=\sum_{\mu\in\mathcal{G}_S\cdot\lambda} m^\mu_\nu(\Phi_L).\label{p}
 \end{equation}
\end{prop}

\begin{example}\
Consider the hybrid character function $\chi_\lambda^L(G_2)(x)$ where $\lambda=2\omega_1+2\omega_2$. If $\omega'_i$ denote the fundamental weights with respect to $\Delta_L$ we have that $N_1\omega+N_2\omega_2=N_1\omega'_1+(N_1+N_2)\omega'_2$ and $r_\beta\cdot(N_1\omega_1+N_2\omega_2)=(N_1+N_2)\omega'_1+N_1\omega'_2$. For notational convenience we represent the dominant integral weights using coordinates with respect to the fundamental weights of the appropriate algebras.  Applying the formula given in \eqref{p} we obtain

{\footnotesize
\begin{align*}
p^{(2,2)}_{(2,2)}(G_2)&= \sum_{(2,4)\in\{(2,4), (4,2)\}} m^{\nu}_{(2,4)}(A_2) =m^{(2,4)}_{(2,4)}(A_2)+m^{(4,2)}_{(2,4)}(A_2)=1+0=1\\
 p^{(2,2)}_{(0,5)}(G_2)&= \sum_{(2,4)\in\{(2,4), (4,2)\}} m^{\nu}_{(0,5)}(A_2)= m^{(2,4)}_{(0,5)}(A_2)+m^{(4,2)}_{(0,5)}(A_2)=1+0=1\\
 p^{(2,2)}_{(2,1)}(G_2)&= \sum_{(2,4)\in\{(2,4), (4,2)\}} m^{\nu}_{(2,3)}(A_2)= m^{(2,4)}_{(2,3)}(A_2)+m^{(4,2)}_{(2,3)}(A_2)=0+1=1\\
 p^{(2,2)}_{(0,4)}(G_2)&= \sum_{(2,4)\in\{(2,4), (4,2)\}} m^{\nu}_{(0,4)}(A_2)= m^{(2,4)}_{(0,4)}(A_2)+m^{(4,2)}_{(0,4)}(A_2)=0+1=1\\
 p^{(2,2)}_{(1,2)}(G_2)&= \sum_{(2,4)\in\{(2,4), (4,2)\}} m^{\nu}_{(1,3)}(A_2)= m^{(2,4)}_{(1,3)}(A_2)+m^{(4,2)}_{(1,3)}(A_2)=2+0=2\\
 p^{(2,2)}_{(1,1)}(G_2)&= \sum_{(2,4)\in\{(2,4), (4,2)\}} m^{\nu}_{(1,2)}(A_2)= m^{(2,4)}_{(1,2)}(A_2)+m^{(4,2)}_{(1,2)}(A_2)=0+2=2\\
 p^{(2,2)}_{(0,2)}(G_2)&= \sum_{(2,4)\in\{(2,4), (4,2)\}} m^{\nu}_{(0,2)}(A_2)= m^{(2,4)}_{(0,2)}(A_2)+m^{(4,2)}_{(0,2)}(A_2)=3+0=3\\
 p^{(2,2)}_{(0,1)}(G_2)&= \sum_{(2,4)\in\{(2,4), (4,2)\}} m^{\nu}_{(0,1)}(A_2)= m^{(2,4)}_{(0,1)}(A_2)+m^{(4,2)}_{(0,1)}(A_2)=0+3=3
\end{align*}
}

These calculations imply that
\begin{align*}
\chi^L_{(2,2)}(x)=&C_{(2,2)}(x)+C_{(0,5)}(x)+C_{(2,1)}(x)+C_{(0,4)}(x)+2C_{(1,2)}(x)\\
                  &+2C_{(1,1)}(x)+3C_{(0,2)}(x)+3C_{(0,1)}(x).
\end{align*}

This formula can be verified by multiplying the right hand side of this expansion by $S^L_{\rho^L}(G_2)(x)$ and comparing with the function $S^L_{\rho^L+\lambda}(G_2)(x)$.
\end{example}

\section{Short root Hybrid Character Functions}

Recall from section 2 that, if $\mathfrak g$ denotes $B_n, C_n, F_4$ or $G_2$, the short root hybrid character functions can be expressed as a linear combination of the orbit functions
$$\chi^S_\lambda(\mathfrak g)(x)= \sum_{\mu\in P^+(\mathfrak g)}q^\lambda_\mu(\mathfrak g)C_\mu(\mathfrak g).$$
Our goal in this section is to determine the coefficients $q^\lambda_\mu(\mathfrak g)$ in terms of the readily calculated multiplicities
$m^\lambda_\mu$.  We follow the same approach that was used for the long root hybrid character functions.

Let $\mathcal{G}_L$ denote the subgroup of the Weyl group $W(\mathfrak{g})$ generated by the reflections $r_\alpha$ where $\alpha$ ranges over the long simple roots of $\mathfrak{g}$. We now list a number of properties of  this subgroup.

\smallskip
\noindent
\textbf{1:} By the definition of the Weyl group $W(\mathfrak{g})$ we have that $\mathcal{G}_L$ is a transversal of the Weyl group $W(\Phi_S)$ of the short root subsystem $\Phi_S$ in the Weyl group $W(\mathfrak{g})$-- i.e. $W(\mathfrak{g})=\bigcup\limits_{\phi\in\mathcal{G}_L}W(\Phi_S)\cdot\phi$.

\smallskip
\noindent
\textbf{2:} By the definition of the sign homomorphism $\sigma^S$ we have that $\sigma^S(\phi)=+1$ for all $\phi\in \mathcal{G}_L$.

\smallskip
\noindent
\textbf{3:} There exists a base of simple roots $\Delta_S$ in $\Phi_S$ such that each $\phi\in \mathcal{G}_L$ permutes these simple roots. In fact we have

\begin{table}[h]
{\footnotesize
\begin{tabular}{|c|c|c|c|}
\hline
\rule{0pt}{10pt}
$\mathfrak g$ & $\Phi_S$ & $\mathcal{G}_L$ & $\Delta_S$ \\ \hline
$B_n$ & $nA_1$  &$\langle r_{\alpha_1},\cdots,r_{\alpha_{n-1}} \rangle$ & $\{e_1,\cdots,e_n\}$
\rule{0pt}{10pt}
\\ \hline
\rule{0pt}{10pt}
$C_n$ & $D_n$ &$ \langle r_{\alpha_n}\rangle$ & $\{e_1-e_2,\cdots, e_{n-1}-e_n, e_{n-1}+e_n\}$
\rule{0pt}{10pt}
\\ \hline
$F_4$ & $D_4$ &$\langle r_\alpha,r_\beta\rangle $ & $\{e_2,\tfrac12(e_1-e_2-e_3-e_4),e_3,e_4\}$  \rule{0pt}{10pt}
\\ \hline
\rule{0pt}{10pt}
$G_2$ & $A_2$ &$\langle r_\alpha\rangle $ & $\{\tfrac13(2e_1-e_2-e_3),\tfrac13(-e_1+2e_2-e_3)\}$
\rule{0pt}{10pt}
\\ \hline
\end{tabular}
}
\end{table}

For each case let $\{\omega''_i\}$ denote the fundamental weights corresponding to the base of simple roots $\Delta_S$. In the table below we provide the expansions for dominant integral weights $\lambda$ of $\mathfrak{g}$ in terms of the $\{\omega''_i\}$ basis. In particular we  observe that each dominant integral weight of $\mathfrak{g}$ is a dominant integral weight of $\Phi_S$. In fact we have

\begin{table}[h]
{\footnotesize
\begin{tabular}{|c|c|c|c|}
\hline
\rule{0pt}{10pt}
$\mathfrak g$ & $\lambda$ & $\Phi_S$ & $\lambda$  \\ \hline
$B_n$ & $\sum\limits_{i=1}^nN_i\omega_i$ & $nA_1$ & $\sum\limits_{i=1}^n 2(N_i+\cdots+\frac{1}{2}N_n)\omega''_i$
\rule{0pt}{10pt}
\\ \hline
\rule{0pt}{10pt}
$C_n$ & $\sum\limits_{i=1}^nN_i\omega_i$ & $D_n$ & $\sum\limits_{i=1}^{n-1}N_i\omega''_i +(N_{n-1}+N_n)\omega''_n$
\rule{0pt}{10pt}
\\ \hline
$F_4$ & $\sum\limits_{i=1}^4N_i\omega_i$ & $D_4$ & $(2N_1+2N_2+N_3)\omega''_1+N_4\omega''_2+(2N_2+N_3)\omega''_3+N_3\omega''_4$  \rule{0pt}{10pt}
\\ \hline
\rule{0pt}{10pt}
$G_2$ & $N_1\omega_1+N_2\omega_2$ & $A_2$ & $(3N_1+N_2)\omega''_1+N_2\omega''_2$
\rule{0pt}{10pt}
\\ \hline
\end{tabular}
\medskip
\caption{{\footnotesize Dominant integral weight $\lambda$ in $\omega$-basis for  the Lie algebras $\mathfrak{g}$ is written in $\omega''_i$-basis for $\Phi_S$.}}
}
\end{table}

\noindent
\textbf{4:} Since each $\phi\in \mathcal{G}_L$ permutes the base of simple roots $\Delta_S$, $\phi$ also permutes the associated fundamental weights $\{\omega''_i\}$ of $\Phi_S$ and hence if $\lambda$ is a dominant integral weight of $\mathfrak{g}$ then $\phi\cdot\lambda$ is also a dominant integral weight of $\Phi_S$. In particular we have that $\phi\cdot \rho^S=\rho^S$.

\smallskip
 Using the properties of $\mathcal{G}_L$ listed above we have that for any dominant weight $\lambda$ of $\mathfrak{g}$

 $$C^S_\lambda(\mathfrak{g})(x)=\sum_{\mu\in\mathcal{G}_L\cdot\lambda}C_\mu(\Phi_S)(x)$$

 $$S^S_{\rho^S+\lambda}(\mathfrak{g})(x)=\sum_{\mu\in\mathcal{G}_L\cdot\lambda}S_{\rho^S+\mu}(\Phi_S)(x)$$

 and finally
 $$\chi^S_\lambda(\mathfrak{g})(x)=\sum_{\mu\in\mathcal{G}_L\cdot\lambda}\chi_\mu(\Phi_S)(x)$$

 Since $\chi_\mu(\Phi_S)(x)=\sum\limits_\nu m^\mu_\nu(\Phi_S)C_\nu(\Phi_S)(x)$ where $m^\mu_\nu(\Phi_S)$ denotes the multiplicity of the dominant integral weight $\nu$ in the simple module of the Lie algebra associated with $\Phi_S$ having highest weight $\mu$. Using this we conclude that

\begin{prop}\

Multiplicity $q^\lambda_\nu(\mathfrak{g})$ of the dominant integral weight $\nu$ in the simple module of the Lie algebra associated with $\Phi_S$ having highest weight $\mu$ for hybrid character $\chi^S_\lambda(\mathfrak g)(x)$ defined by formula \eqref{Scharacter} has a  following form
 \begin{equation}\label{q}q^\lambda_\nu(\mathfrak{g})=\sum_{\mu\in\mathcal{G}_L\cdot\lambda} m^\mu_\nu(\Phi_S).\end{equation}
\end{prop}

\begin{example}\

Consider the hybrid character function $\chi_\lambda^S(G_2)(x)$ where $\lambda=2\omega_1+\omega_2$. If $\omega''_i$ denote the fundamental weights with respect to $\Delta_S$ we have  that $N_1\omega+N_2\omega_2=(3N_1+N_2)\omega''_1+N_2\omega''_2$ and $r_\alpha\cdot(N_1\omega_1+N_2\omega_2)=N_2\omega''_1+(N_3+N_2)\omega''_2$. For notational convenience we represent the dominant integral weights using coordinates with respect to the fundamental weights of the appropriate algebras.  Applying the formula given in \eqref{q} we obtain

{\footnotesize
\begin{eqnarray*}
 q_{(2,1)}^{(2,1)}(G_2) = \sum_{\nu\in\{(7,1), (1,7)\}} m^{\nu}_{(7,1)}(A_2) = m^{(7,1)}_{(7,1)}(A_2)+m^{(1,7)}_{(7,1)}(A_2)= 1+0=1\\
 q_{(1,2)}^{(2,1)}(G_2) = \sum_{\nu\in\{(7,1), (1,7)\}} m^{\nu}_{(5,2)}(A_2) = m^{(7,1)}_{(5,2)}(A_2)+m^{(1,7)}_{(5,2)}(A_2)=0+1=1\\
 q_{(2,0)}^{(2,1)}(G_2) = \sum_{\nu\in\{(7,1), (1,7)\}} m^{\nu}_{(6,0)}(A_2) = m^{(7,1)}_{(6,0)}(A_2)+m^{(1,7)}_{(6,0)}(A_2)=0+2=2\\
 q_{(0,3)}^{(2,1)}(G_2) = \sum_{\nu\in\{(7,1), (1,7)\}} m^{\nu}_{(3,3)}(A_2) = m^{(7,1)}_{(3,3)}(A_2)+m^{(1,7)}_{(3,3)}(A_2)= 1+1=2\\
 q_{(1,1)}^{(2,1)}(G_2) = \sum_{\nu\in\{(7,1), (1,7)\}} m^{\nu}_{(4,1)}(A_2) = m^{(7,1)}_{(4,1)}(A_2)+m^{(1,7)}_{(4,1)}(A_2)=1+2=3\\
 q_{(0,2)}^{(2,1)}(G_2) = \sum_{\nu\in\{(7,1), (1,7)\}} m^{\nu}_{(2,2)}(A_2) = m^{(7,1)}_{(2,2)}(A_2)+m^{(1,7)}_{(2,2)}(A_2)=2+2=4\\
 q_{(1,0)}^{(2,1)}(G_2) = \sum_{\nu\in\{(7,1), (1,7)\}} m^{\nu}_{(3,0)}(A_2) = m^{(7,1)}_{(3,0)}(A_2)+m^{(1,7)}_{(3,0)}(A_2)= 2+2=4\\
 q_{(0,1)}^{(2,1)}(G_2) = \sum_{\nu\in\{(7,1), (1,7)\}} m^{\nu}_{(1,1)}(A_2) = m^{(7,1)}_{(1,1)}(A_2)+m^{(1,7)}_{(1,1)}(A_2)=2+2=4\\
 q_{(0,0)}^{(2,1)}(G_2) = \sum_{\nu\in\{(7,1), (1,7)\}} m^{\nu}_{(0,0)}(A_2) = m^{(7,1)}_{(0,0)}(A_2)+m^{(1,7)}_{(0,0)}(A_2)=2+2=4
\end{eqnarray*}
}

These calculation then give

\begin{align*}
  \chi^S_{(2,1)}(x)=&C_{(2,1)}(x)+C_{(1,2)}(x)+2C_{(2,0)}(x)+2C_{(0,3)}(x)+3C_{(1,1)}(x)\\
  &+4C_{(0,2)}(x)+4C_{(1,0)}(x) +4C_{(0,1)}(x)+4C_{(0,0)}(x)
\end{align*}

This decomposition can be verified by multiplying the right hand side of the expansion by $S^S_{\rho^S}(G_2)(x)$ and observing that this product yields $S^S_{\rho^S+\lambda}(G_2)(x)$.

\end{example}

\section{Dominant weight multiplicities for the hybrid characters}

In the table below we summarize the dominant weight multiplicities for each of the hybrid characters.
\begin{table}[h]
{\footnotesize
\begin{tabular}{|c|c|c|}
\hline
\rule{0pt}{10pt}
$\mathfrak g$ & $p^\lambda_\mu(\mathfrak g)$ & $q^\lambda_\mu(\mathfrak g)$ \\ \hline
$B_n$ & $\sum\limits_{\nu\in\{\lambda,r_{\alpha_n}\cdot\lambda\}}m^\nu_\mu(D_n)$  & $\sum\limits_{\nu\in S_n\cdot \lambda} m^\nu_\mu(nA_1)$
\rule{0pt}{10pt}
\\ \hline
\rule{0pt}{10pt}
$C_n$ & $\sum\limits_{\nu\in S_n\cdot \lambda} m^\nu_\mu(nA_1)$ & $\sum\limits_{\nu\in\{\lambda,r_{\alpha_n}\cdot\lambda\}}m^\nu_\mu(D_n)$
\rule{0pt}{10pt}
\\ \hline
$F_4$ & $\sum\limits_{\nu\in \mathcal{G}_S\cdot\lambda}m^\nu_\mu(D_4)$ & $\sum\limits_{\nu\in{\mathcal{G}_L}\cdot\lambda}m^\nu_\mu(D_4)$  \rule{0pt}{10pt}
\\ \hline
\rule{0pt}{10pt}
$G_2$ & $\sum\limits_{\nu\in\{\lambda, r_{\beta}\cdot\lambda\}}m^\nu_\mu(A_2)$ & $\sum\limits_{\nu\in\{\lambda, r_{\alpha}\cdot\lambda\}}m^\nu_\mu(A_2)$
\rule{0pt}{10pt}
\\ \hline
\end{tabular}
\medskip
\caption{{\footnotesize Dominant weight multiplicities $p^\lambda_\mu(\mathfrak g)$ and $q^\lambda_\mu(\mathfrak g)$ for hybrid characters $\chi^L(\mathfrak g)$ and $ \chi^S(\mathfrak g)$ of simple Lie algebra $\mathfrak g$. They are expressed in terms of multiplicities of dominant weights for certain reducible representations of the Lie algebras of long and short roots subsystems of $\mathfrak g$. In the line for the algebra $F_4$, $\mathcal{G}_L$ denotes the subalgebra of $W(F_4)$ generated by $\langle r_\alpha, r_\beta\rangle$ and $\mathcal{G}_S$ denotes the subalgebra of $W(F_4)$ generated by $\langle r_\gamma, r_\delta\rangle$.}}}
\end{table}


\section{Interpretations and Weyl formulas for $\chi^L_\lambda(\mathfrak{g})(0)$ and $\chi^S_\lambda(\mathfrak{g})(0)$}\

In the case of the irreducible character function $\chi_\lambda(\mathfrak{g})(x)$ it is well known that $\chi_\lambda(\mathfrak{g})(0)$ is equal to the dimension of the simple $\mathfrak{g}$ module having highest weight $\lambda$. In addition we also have the Weyl formula

$$\chi_\lambda(\mathfrak{g})(0)= \frac{\prod (\lambda+\rho,\alpha)}{\prod(\rho,\alpha)}$$
where the products range over the positive roots of the $\mathfrak{g}$ root system.

In this section we provide ``dimensional" interpretations and Weyl type formulas for  $\chi^L_\lambda(\mathfrak{g})(0)$ and $\chi^S_\lambda(\mathfrak{g})(0)$. We will start with the long root hybrid character functions.


Recall that for the long root hybrid character functions we have
$$\chi^L_\lambda(\mathfrak{g})(x)=\sum_{\mu\in\mathcal{G}_S\cdot\lambda}\chi_\mu(\Phi_L)(x)$$
therefore
$$\chi^L_\lambda(\mathfrak{g})(0)=\sum_{\mu\in\mathcal{G}_S\cdot\lambda}\chi_\mu(\Phi_L)(0)$$

Since the dimension of the simple $\Phi_L$ module with highest weight $\lambda$, denoted  $L_\lambda(\Phi_L)$,  is equal to the dimension of the simple $\Phi_L$ module with highest weight $\phi\cdot \lambda$ we conclude that
\begin{equation}\chi^L_\lambda(\mathfrak{g})(0)=|\mathcal{G}_S\cdot\lambda|\dim L_\lambda(\Phi_L)\end{equation}

It follows immediately then that
\begin{equation}\chi^L_\lambda(B_n)(0)=|\mathcal{G}_S\cdot \lambda|\frac{\prod (\lambda+\rho^L(\mathfrak{g}),\alpha)}{\prod (\rho^L(\mathfrak{g}),\alpha)}\end{equation}
where the products range over the positive long roots $\alpha$ of $\mathfrak{g}$.


We have analogous results for the short root characters functions. In fact, recall that for the short root hybrid character functions we have
$$\chi^S_\lambda(\mathfrak{g})(x)=\sum_{\mu\in\mathcal{G}_L\cdot\lambda}\chi_\mu(\Phi_S)(x)$$
therefore
$$\chi^S_\lambda(\mathfrak{g})(0)=\sum_{\mu\in\mathcal{G}_L\cdot\lambda}\chi_\mu(\Phi_S)(0)$$

Since the dimension of the simple $\Phi_S$ module with highest weight $\lambda$, denoted  $L_\lambda(\Phi_S)$,  is equal to the dimension of the simple $\Phi_S$ module with highest weight $\phi\cdot \lambda$ for any $\phi\in\mathcal{G}_L$, we conclude that
\begin{equation}\chi^S_\lambda(\mathfrak{g})(0)=|\mathcal{G}_L\cdot\lambda|\dim L_\lambda(\Phi_S)\end{equation}

It follows immediately then that
\begin{equation}\chi^S_\lambda(\mathfrak{g})(0)=|\mathcal{G}_L\cdot \lambda|\frac{\prod (\lambda+\rho^S(\mathfrak{g}),\alpha)}{\prod (\rho^S(\mathfrak{g}),\alpha)}\end{equation}
where the products range over the positive short roots $\alpha$ of $\mathfrak{g}$.

\section{Concluding remarks}
With the determination of the dominant weight multiplicities for the hybrid characters, all but the most curious property of these functions have been revealed. It remains to be seen whether these functions can be interpreted as characters of some algebraic structures.

Combining pairs of the functions $C$, $S$ forms multivariable generalizations of the common exponential functions. Combining pairs of $C$, $S$ $S^L$, and $S^S$ functions,  the number of families of $E$-functions is increased from one to six families \cite{HHP}.

Discretization of the families of these functions hinges on discretization of the $C$-functions \cite{HMP1,HMP2}, which is now more that 20 years old \cite{MP84}.

Another class of hybrid characters arises for simple Lie algebras with non-trivial automorphism of their Coxeter-Dynkin diagram. They will be defined and studied elsewhere.

\subsection*{Acknowledgements}\

We gratefully acknowledge the support of this work by the Natural Sciences and Engineering Research Council of Canada and by the MIND Research Institute of Irvine, California. M.S. would like to express her gratitude to the Centre de recherches math\'ematiques, Universit\'e de Montr\'eal, for the hospitality extended to her during her postdoctoral fellowship. M.S. is also grateful to MITACS and to OODA Technologies Inc. for partial support.


\end{document}